\documentclass[preprint]{aastex} 
\usepackage{graphicx}
\usepackage{txfonts}
\usepackage{natbib} 
\usepackage{url} 
\usepackage{fullpage}
\bibpunct{(}{)}{;}{a}{}{,} 
\begin{document}
%
   \title{Extreme Ultraviolet Solar Irradiance during the rising phase of solar cycle 24 observed by PROBA2/LYRA}


   \author{M. Kretzschmar$^{1,2}$, I.E. Dammasch$^{3}$, M. Dominique$^{3}$, J. Zender$^{4}$, G. Cessateur$^{5}$, E. D'Huys$^{3}$}
   \affil{$^{1}$Royal Observatory of Belgium / Solar Terrestrial Center of Excellence, av. circulaire 3, 1180 Brussels, Belgium\\}
   \affil{$^{2}$LPC2E, UMR 6115 CNRS and University of Orl\'eans, 3a av. de la recherche scientifique, 45071 Orl\'eans, France\\}
   \affil{$^{3}$Royal Observatory of Belgium / Solar Influence Data Analysis Center, av. circulaire 3, 1180 Brussels, Belgium\\}
   \affil{$^{4}$ESA Research and Scientific Support Department, ESTEC, NL-2200 AG Noordwijk, The Netherlands}
   \affil{$^{5}$Physikalisch-Meteorologisches Observatorium Davos and World Radiation Center (PMOD/WRC), Dorfstrasse 33, 7260 Davos Dorf, Switzerland}
    \email{matthieu.kretzschmar@cnrs-orleans.fr}



 
\begin{abstract}
The Large Yield Radiometer (LYRA) is a radiometer that has monitored the solar irradiance at high cadence and in four pass bands since January 2010. Both the instrument and its spacecraft, PROBA2 (Project for On-Board Autonomy), have several innovative features for space instrumentation, which makes the data reduction necessary to retrieve the long term variations of solar irradiance more complex than for a fully optimized solar physics mission. In this paper, we describe how we compute the long term time series of the two extreme ultraviolet irradiance channels of LYRA, and compare the results with SDO/EVE. We find that the solar EUV irradiance has increased by a factor 2 since the last solar minimum (between solar cycles 23 and 24), which agrees reasonably well with the EVE observations.
  \end{abstract}
%
%
\section{Introduction}
Solar extreme ultraviolet (EUV) irradiance affects the Earth's upper atmosphere by heating, exciting, dissociating and ionizing its main constituents: O$_{2}$, N$_{2}$, and O. It is responsible for the diurnal ionosphere and influences important parameters for space weather, such as the total electron content (TEC), important for telecommunications, and the thermosphere density, important for satellite drag \citep{Solomon:2010qy,2011GeoRL..3819102D}. \\
Being completely absorbed by the Earth's atmosphere, measurements of solar EUV irradiance started with the space age; consequently, the knowledge of the absolute value of the incoming EUV flux as well as its variations with solar activity is challenged by the available technology and how it reacts to space environment. In particular, degradation of the instrument sensitivity caused by a combination of contaminants and long exposure to the sunlight makes it difficult to truly assess the irradiance variations on the mid and long term. Until 2002 and the observations by the Solar EUV Experiment (SEE) aboard the Thermosphere Ionosphere Mesosphere Energetics and Dynamics (TIMED) mission \citep{Woods:2005aa}, the monitoring of the solar EUV flux has been scarce and modelers of the thermosphere/ionosphere system have relied on proxy-based models. These models, based on solar proxies like the 10.7cm radio flux, have been very useful but their precision has proven to be insufficient for space weather operations. General reviews on the EUV irradiance can be found in \cite{Lean:1987ab,1993JGR....9818879T,2008AnGeo..26..269L,Lean:2011fk}.

The long term variations of the EUV irradiance are important to monitor for at least three reasons: their impact on Earth and human activities in space, as proxy of the changes in the solar upper atmosphere, and to understand its contribution to the variation of the solar total irradiance (even if minor). However, they are difficult to determine because of limiting lifetime of the missions and the aging of the instrumentation in space. The only data set that covers one full solar cycle is the EUV flux integrated from 0.1nm to 50nm and from 26nm to 34nm measured by the Solar EUV Monitor (SEM) on the Solar and Heliospheric Observatory (SOHO). Recent analysis of these data has shown that long term variations of the EUV flux do impact the long term variations of thermospheric density and temperature \citep{2011JGRA..11600H07S}, which makes their knowledge compulsory for spacecraft operations. Ideally, forecasts of the EUV flux several years in advance is required. The monitoring of the EUV emission, which originates in the solar upper atmosphere, reveals how the solar corona evolves over long time scales. Recently, two missions with EUV irradiance instruments onboard have been launched and are presently monitoring the rise of solar cycle 24. The first one is the PROBA2 mission (ESA) that includes the Large Yield RAdiometer (LYRA) \citep{Hochedez:2006aa, Dominique:2012aa}. LYRA monitors the solar irradiance in four pass-bands and has been operating since early January 2010. The second mission is the NASA Solar Dynamics Observatory (SDO) with the EUV Variability Experiment (EVE) \citep{Woods:2010zr}. EVE is a combination of several instruments that can monitors the solar irradiance spectrum from 6nm to 105nm with a spectral resolution of 0.1nm, and the irradiance in larger pass-bands between 0.1 and 17 nm and around the Lyman-$\alpha$ line at 121.6nm. Unfortunately, up to now, spectral measurements are only available up to 65nm. 

The PROBA2 mission has been conceived to be a technology demonstrator and has several innovative features onboard. The LYRA instrument itself is the first instrument in space that has diamond detectors. The combination of the spacecraft and instrument innovations and specificities leads to instrumental artifacts in the data, which prevent easily retrieval of representative values of solar irradiance for medium and long term studies, e.g., daily irradiance value. In this paper, we first present in details these instrumental artifacts and derive daily irradiance time series over the lifetime of the mission (Sec.\ref{sec_data}). This new LYRA data product is named \emph{LYRA$\_$yyyymmdd$\_$lev3d$\_$merged.txt}, where yyyymmdd stands for the date of the last available data and lev3d indicates the daily time step (level2 is full resolution and standard level 3 is 1 minute resolution); it will be made available at \url{http://proba2.sidc.be/index.html/Data-download/article/swap-lyra-data-distribution}. We restrict ourself to the two EUV channels of LYRA (channel 3 and 4), covering wavelengths from soft X-ray (SXR) to 80nm. We next compare these time series to other observations and solar activity indices (Sec.\ref{sec_results}) to gain information on the solar variability in the EUV in the rising phase of solar cycle 24.  
%
%
%
\section{The Large-Yield RAdiometer onboard PROBA2 \label{sec_data}}
\subsection{Dark current correction, degradation, and calibration \label{sec_lyracal}}
\begin{figure}[t]    
   \centerline{\includegraphics[width=0.7\textwidth]{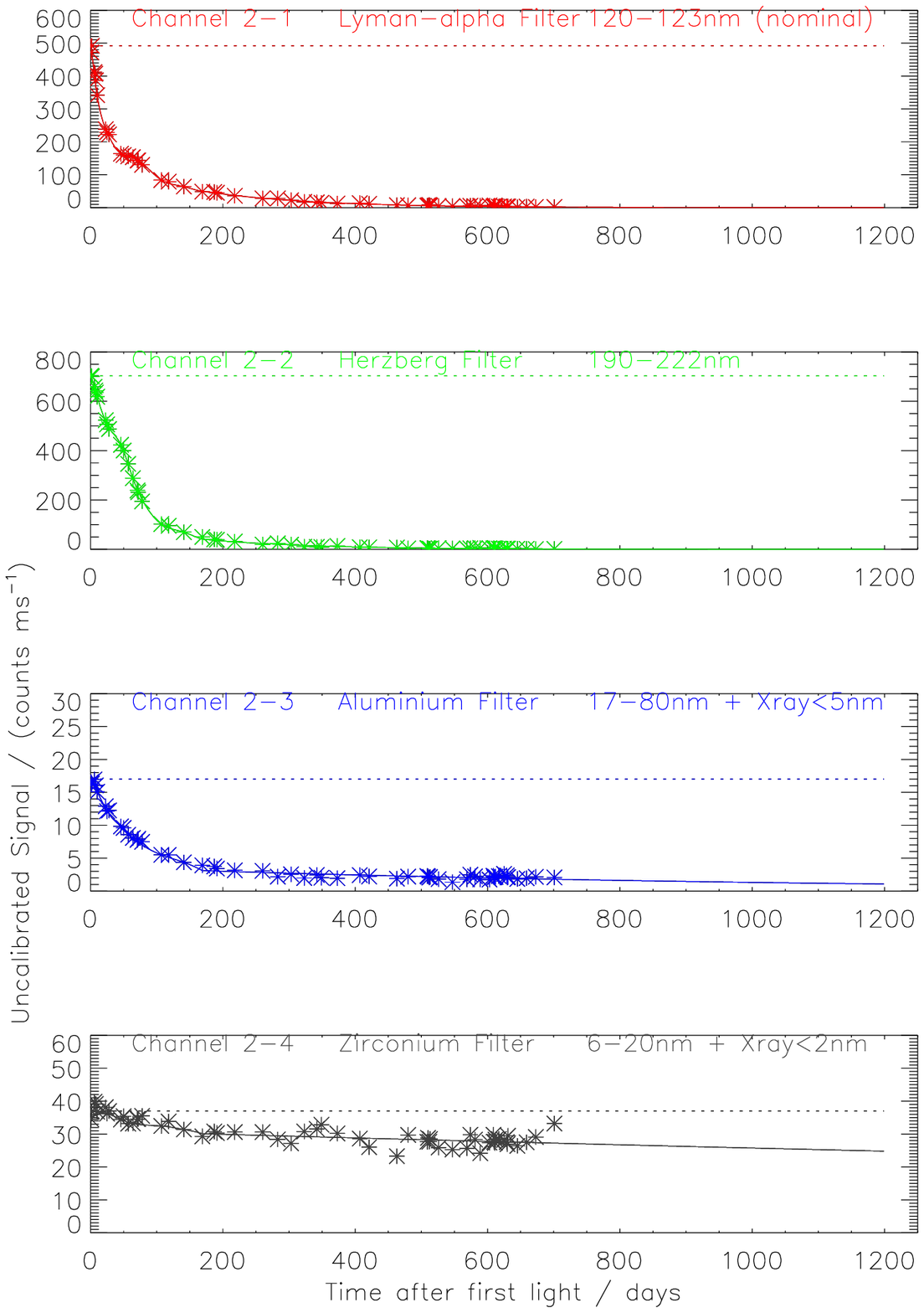}
             }
              \caption{\textbf{Degradation of the 4 channels of LYRA unit 2} From top to bottom, channel 1 (Lyman-$\alpha$), channel 2 (Herzberg), channel 3 (Aluminium filter), and channel 4 (Zirconium filter). }
   \label{fig_deg}
   \end{figure}
 \begin{figure}[h]    
   \centerline{\includegraphics[width=0.49\textwidth,clip=]{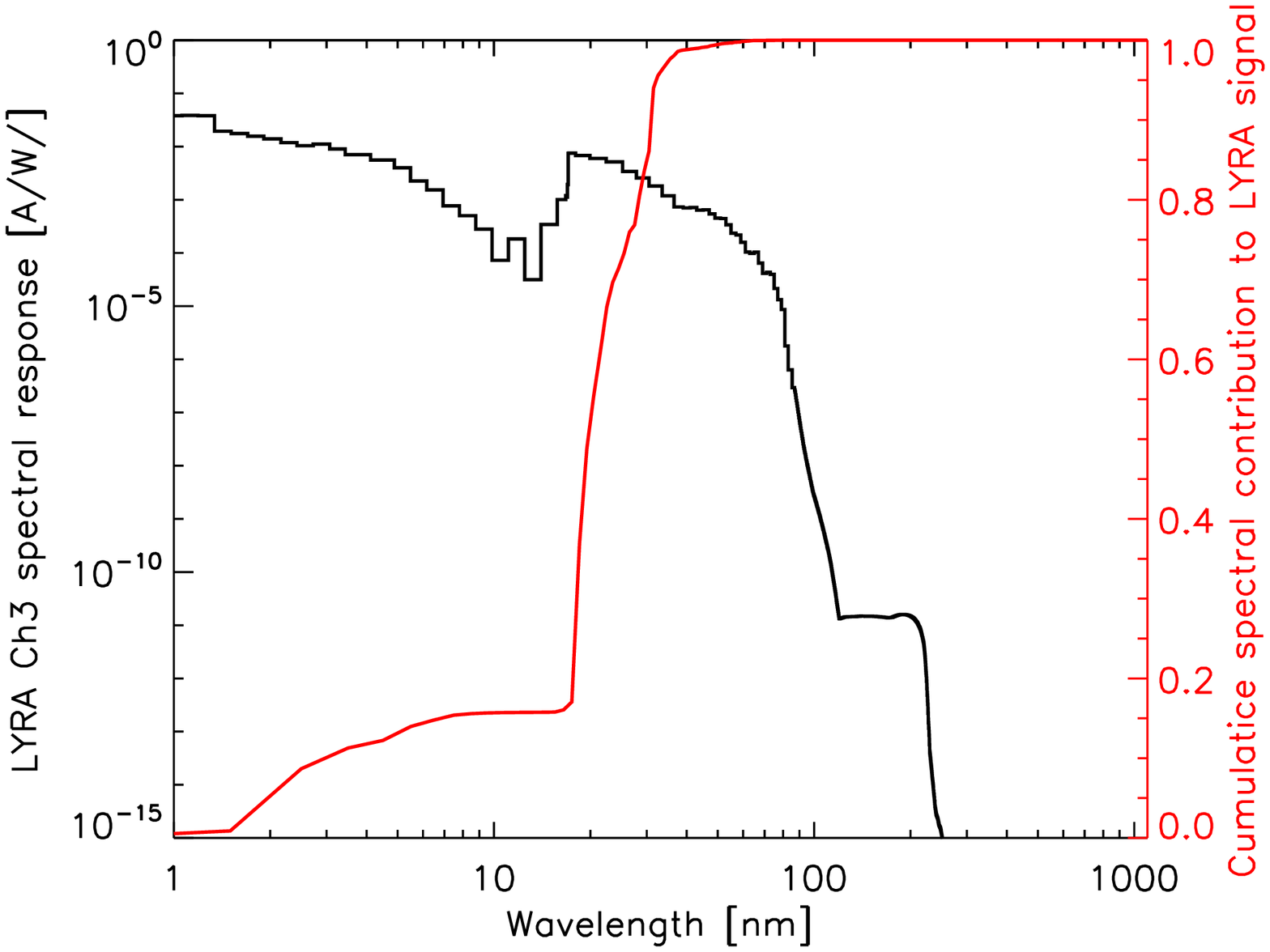}\includegraphics[width=0.49\textwidth,clip=,]{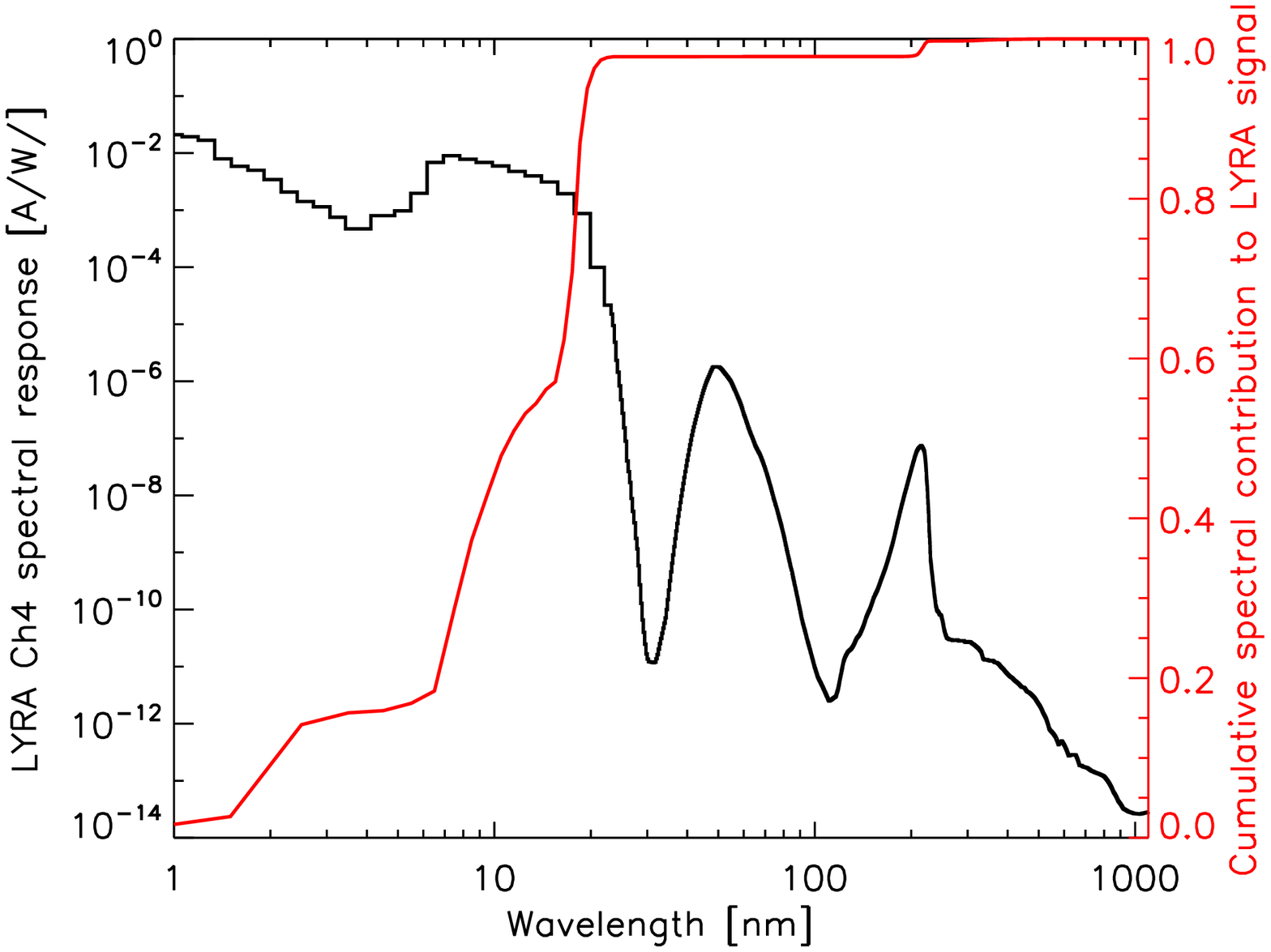}
             }
              \caption{\textbf{Spectral response of LYRA unit 2 channel 3 (left) and 4 (right).} The cumulative spectral contribution has been determined by multiplication of the spectral response by the solar spectrum for 1 February 2010 -quiet solar conditions- as deduced from TIMED/SEE and two instruments aboard the spacecraft SOlar Radiation and Climate Experiment (	): the spectral irradiance monitor (SIM) and the SOlar Stellar Irradiance Comparison Experiment (SOLSTICE) (see text for details).}
   \label{fig_resp}
   \end{figure}
The LYRA instrument is a small radiometer (315 x 92.5 x 222 mm) composed of three redundant units, all three hosting the same four broad-bandpass channels. One of these units is used continuously, while the two others are used for punctual observation and calibration campaigns, respectively, and are experiencing a much lower degradation. Those three units are similar in bandpasses, but involve different combinations of filters and detectors. Besides the use of the innovative diamond detectors (both EUV channels of unit 2 are made of MSM diamond detectors), the main characteristics of the instrument is its high acquisition rate: 20Hz (nominal).
%

PROBA2 revolves on a helio-synchronous (polar) orbit at about 720 km altitude. The orbit is oriented in such a way that the instrument faces the Sun most of the time. Short eclipses of maximum 20 min are nevertheless experienced in each orbit (i.e., every 100 min) during the annual occultation season, which is described in section \ref{ch:additional_corrections}.  
\section{Generation of daily irradiance time series}
The LYRA level 1 data products are uncorrected count rates. The level 2 (full cadence) and level 3 (1-minute average) data are corrected for dark current, degradation, and converted into physical units (W/m$^{2}$). \\
Dark current is estimated from pre-launch measurements and onboard observations with the cover closed \citep{2009A&A...508.1085B}. It is strongly dependent on temperature and varies in phase with the orbit; other variations caused by occultation, seasonal effects, or local conditions of the spacecraft must be removed too. The LYRA onboard temperature is registered near the detector at three different locations. 

Degradation is one of the main challenges for space measurements of solar ultraviolet radiation. LYRA has experienced very large degradation in its two UV channels (around Lyman-$\alpha$ and 210nm), and significant degradation in the two extreme UV (EUV) channels.
The current degradation estimate is made by comparing the nominal unit (unit 2, whose cover has been open almost continuously for two years) with the back-up unit 3 (that was opened in total for about 6 days by the end of 2011). The last unit, unit 1, opened for about two days only; this will be used in future versions of the degradation correction. The signal observed in the channel 4 of unit 3 does not show any degradation and has been used as reference for estimating the degradation in the channels of the nominal unit 2. The degradation curves are fitted by an exponential function and are added to the actual observed count rate (after dark current subtraction); use of an additive correction was chosen since the degradation is thought to occur mainly in the (longer wavelength) EUV part of the LYRA channels, while the SXR part responsible for the flares would remain unaffected. Fig.\ref{fig_deg} shows the current estimate of the degradation in each of the four channels of the nominal unit of LYRA. In this paper, we concentrate on the time series of the channel 3 and 4 of LYRA unit 2. 

The calibration of LYRA is based on pre-launch measurements of the spectral response of the filter/detector combination. Fig.\ref{fig_resp} shows the spectral response of channels 3 and 4. It is clear that both channels have both an EUV contribution and an SXR contribution. Indeed the nominal passbands are defined as follow: from 17 to 80nm and below 5nm for channel 3 and from 6 to 20nm and below 2nm for channel 4. In the following, we refer to the EUV contribution to channel 3 and 4 as the contribution coming from wavelengths above 17nm and 6nm, respectively; the contribution coming from shorter wavelengths is named SXR. On first light, the difference averaged over the three units between observed and simulated (using TIMED/SEE and SORCE observations) count rate was 13\% for channel 3 and 9\% for channel 4. 
\subsection{Correction of spacecraft, instrument and environmental effects}\label{ch:additional_corrections}
 \begin{figure}[t]    
   \centerline{\includegraphics[width=0.49\textwidth,clip=,bb=1 170 550 580]{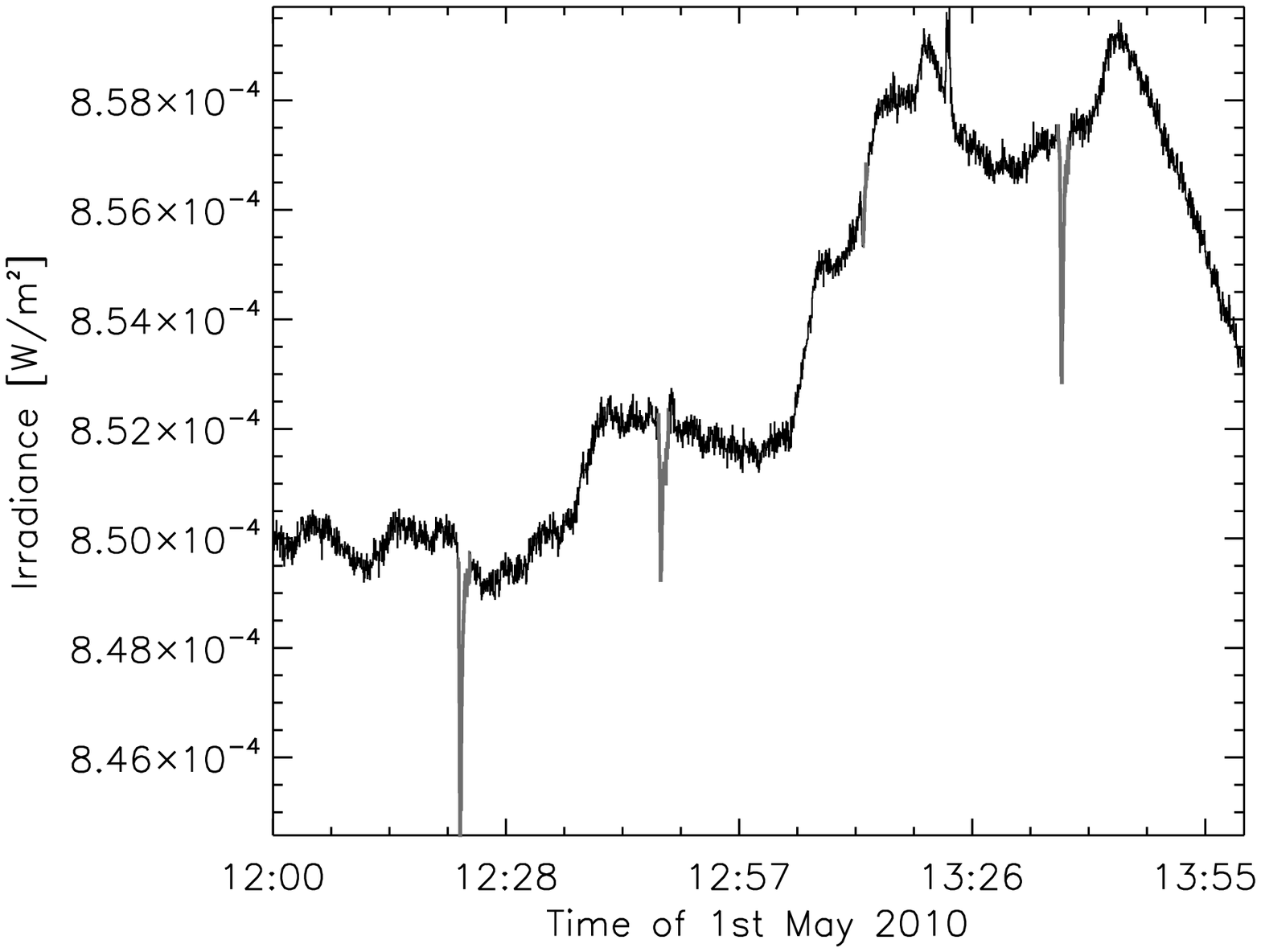}\includegraphics[width=0.49\textwidth,clip=,bb=15 170 570 580]{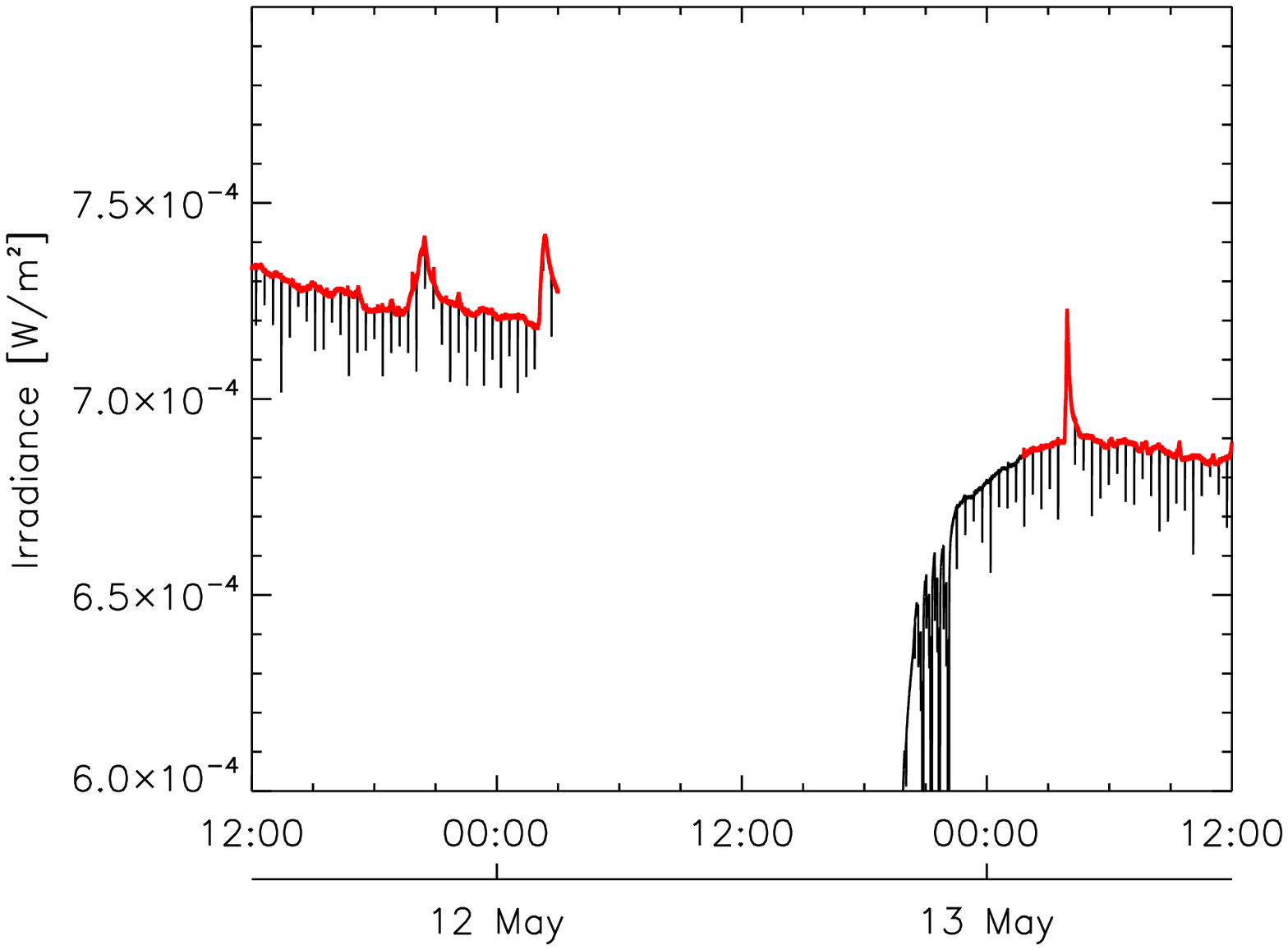}
             }
              \caption{\emph{Left panel:} Effects of the Large Angle Rotations of PROBA2 on the channel 4 time series. Periods overlaid in grey indicate data to be removed and have been determined using the housekeeping information plus an algorithm to detect the dips. \emph{Right panel:}In black, the original level 2 data. In red, the data corrected for LARs and slow stabilization after cover opening. }
   \label{fig_lar}
   \end{figure}
The calibrated level 2 or level 3 data still suffer from several instrumental effects:
\begin{itemize}
\item \textbf{Large angle rotations (LARs).}
Approximately every 25 min., the spacecraft is rotating by 90 degrees around the spacecraft-Sun axis. This movement provokes sharp peaks and dips in channel 3 and 4 that are at present not easily correctable. We used the spacecraft information to localize these LARs, and then performed some analysis to determine the period over which the irradiance values are affected. The procedure first identifies the reversed peak in channel 4 (LARs appear mostly as dips in channel 4) in the vicinity of the LAR time as indicated by the spacecraft. The start of the peak is defined as the last point having a positive or null slope minus 6 seconds. The end of the dip is defined as the first point whose value is within 0.015\% of the start point, within the limit of 90s after the reversed peak. In cases where the procedure based on the localization of the reverse peak failed, we removed the data between 45s before and 33s after the time indicated by the ancillary parameters; these values have been determined from the analysis of all detected LARs in February 2010. Results of this analysis are shown in Fig.\ref{fig_lar}.
%
%
%
\item \textbf{Stabilization after cover closing/opening.} MSM  diamond detectors, that constitute the channels 1, 3 and 4 of the nominal unit, need to have been exposed a certain time before reacting nominally. This is due to the presence of surface defects on the detectors, that trap the photoelectrons before they are collected by the electrodes. This phenomenon affects the measurements until all the traps have been "filled". The reverse effect is observed after closing the covers. All the trapped electrons are progressively released and collected by the electrodes. The consequence is that the observed flux takes some time before reaching its actual level. Such an effect is not observed with the PIN diamond detectors where the electrons are produced deeper, in the bulk of the detector. The effects of this trapping can be observed on the right panel of Fig.\ref{fig_lar} which shows the slow stabilization of the flux level after the cover was closed and opened again on May 12 at 19:30. Empirical observations have shown that removing the first 6 hours of acquisition after cover opening is reasonable. This is how we corrected for this effect.

%
%
%
 \begin{figure}[t]    
   \centerline{\includegraphics[width=0.6\textwidth,bb=0 170 570 580,clip=]{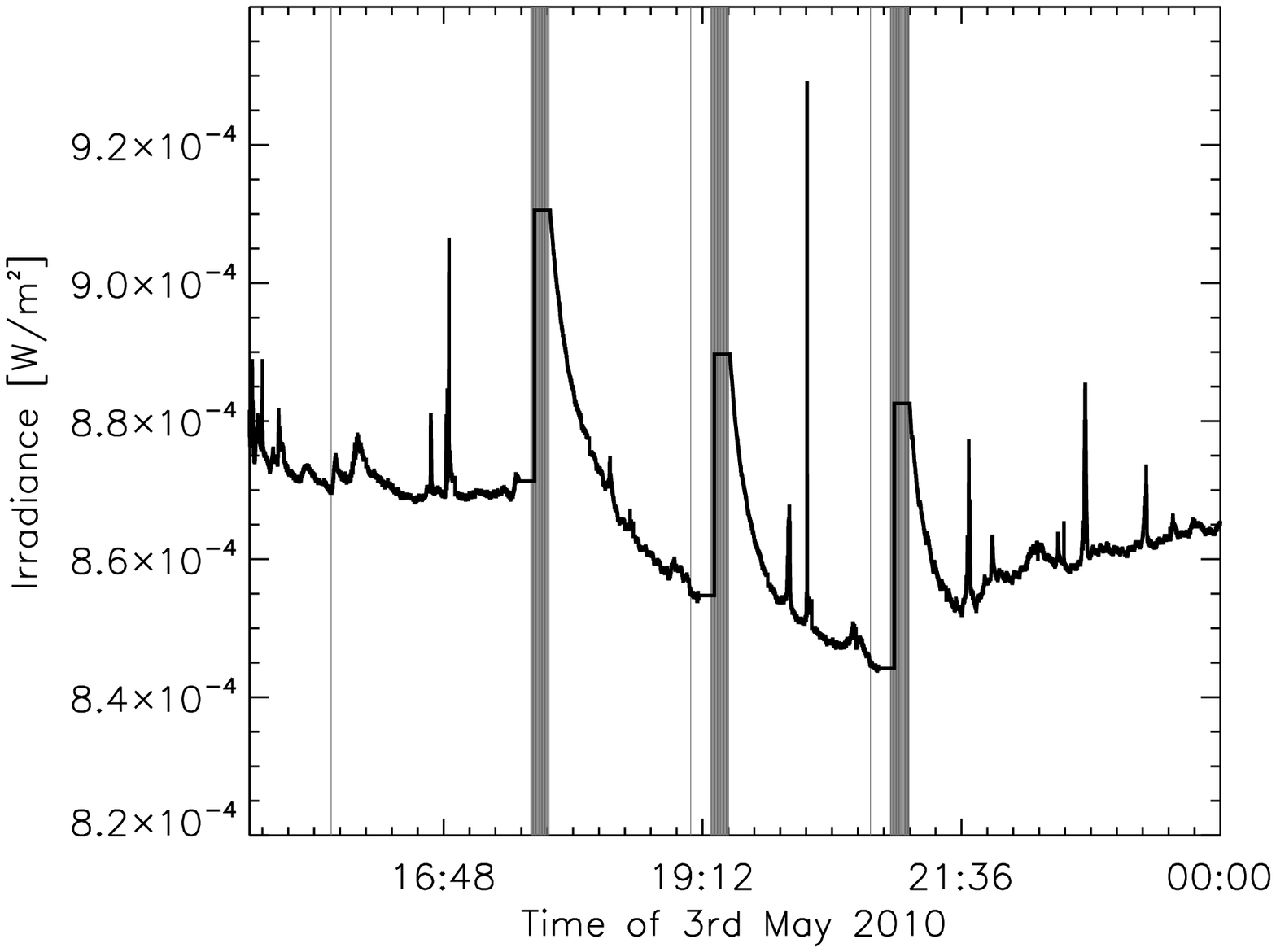}
             }
              \caption{\textbf{Effects of the reboots on the Lyra channel 4 time series.} Vertical grey lines indicate time of reboots. Thick lines are actually showing several reboots close in time.}
   \label{fig_vfc}
   \end{figure}
\item \textbf{Peak after instrument switch on or reload.} Right after the switch on or the reload of the instrument, the first samples acquired are removed since they are likely to show undesired behaviors. However, as can be seen in Fig.\ref{fig_vfc} the effect of these operations persists for some time in some of the LYRA channels; accordingly, the data acquired during the next hour after a reload were removed for the computation of the daily value.
The number of those reloads is now limited but this operation was needed at the beginning of the mission, in order to (de)activate units, and this effect was observed anytime a back-up unit was used. 
%
%
\end{itemize} 
Other non-instrumental effects must be removed from the data
\begin{itemize}
\item \textbf{Eclipses and occultations.} Here the term eclipse designates the passage of the moon in front of the Sun as seen from the PROBA2 spacecraft; it occurs about three times a year and can be predicted easily. Occultations indicate the occultation of the Sun by the Earth or its atmosphere from the spacecraft point of view. PROBA2 has one occultation season per year, from approximately November to February. Channels 3 and 4 of LYRA, which monitors the incoming EUV flux, clearly show the EUV flux absorption by the Earth's upper atmosphere followed by the occultation by the Earth when the LYRA signal reaches its dark current level. These data are still present in the LYRA level 1, level 2 (usually 50 ms cadence), and level 3 (1 minute average) files but must be removed to compute daily values. To do so, we simply used the spacecraft position information and kept only data for which the line of sight from the spacecraft to the Sun is not crossing the Earth's upper atmosphere.
\item \textbf{South Atlantic Anomaly.} It is well known that spacecraft flying over the south atlantic encounter high levels of radiation in a localized region named the South Atlantic Anomaly (SAA). Channel 3 and 4 of the nominal unit of LYRA used MSM diamond detectors that are much more resistant to energetic radiation than silicon detectors \citep[see][]{Dominique:2012aa} so that the SSA is usually not seen in these data. Therefore, we did not introduce a specific correction for these data.
\end{itemize} 
\hspace{0.8cm} Finally, we normalized the observed flux value to 1 astronomical unit (AU).
The data corrected for these effects still have some instrumental features of unclear origin. We additionally performed some filtering in order to remove outliers that cannot be deduced from the LYRA metadata. To this aim, we identified the bad measurements in the LYRA channel 1 (Lyman-$\alpha$) time series since it has less sensitivity to solar variability as a consequence of degradation. This filtering was done in two steps: using 3 seconds rebinned LYRA irradiance, we first removed the data points for which the channel 1 values deviate from the hourly median by more than 0.25\%. Next, we re-computed the daily median without these points, and removed the data points for which the channel 1 values deviates by more than  0.1\% of the daily median. With this correction, it is actually possible that valid measurements were discarded in computing the daily irradiance values; however, the impact on the daily values is negligible as it has been checked by testing different threshold. Let us note that we have also derived the daily irradiance values after having removed the data obtained during flares (as deduced from the GOES SXR start and end times of the events) in the original high cadence data. However, this does not lead to any significant changes in the mid- and long- term behavior of the daily irradiance, suggesting that the SXR contribution to the daily value is mostly coming from active region emission.

These daily LYRA irradiances are to be considered as version 1 of this data product. The catalog of events impacting the LYRA data, whether they are from instrumental origins or natural phenomena, is in constant evolution and new versions of the LYRA daily irradiance values based on the described scheme will be released in the future. 
%
%
\section{Results and discussions\label{sec_results}}
%
%
 \begin{figure}[t]    
   \centerline{\includegraphics[width=0.49\textwidth,bb=40 190 600 600,clip=]{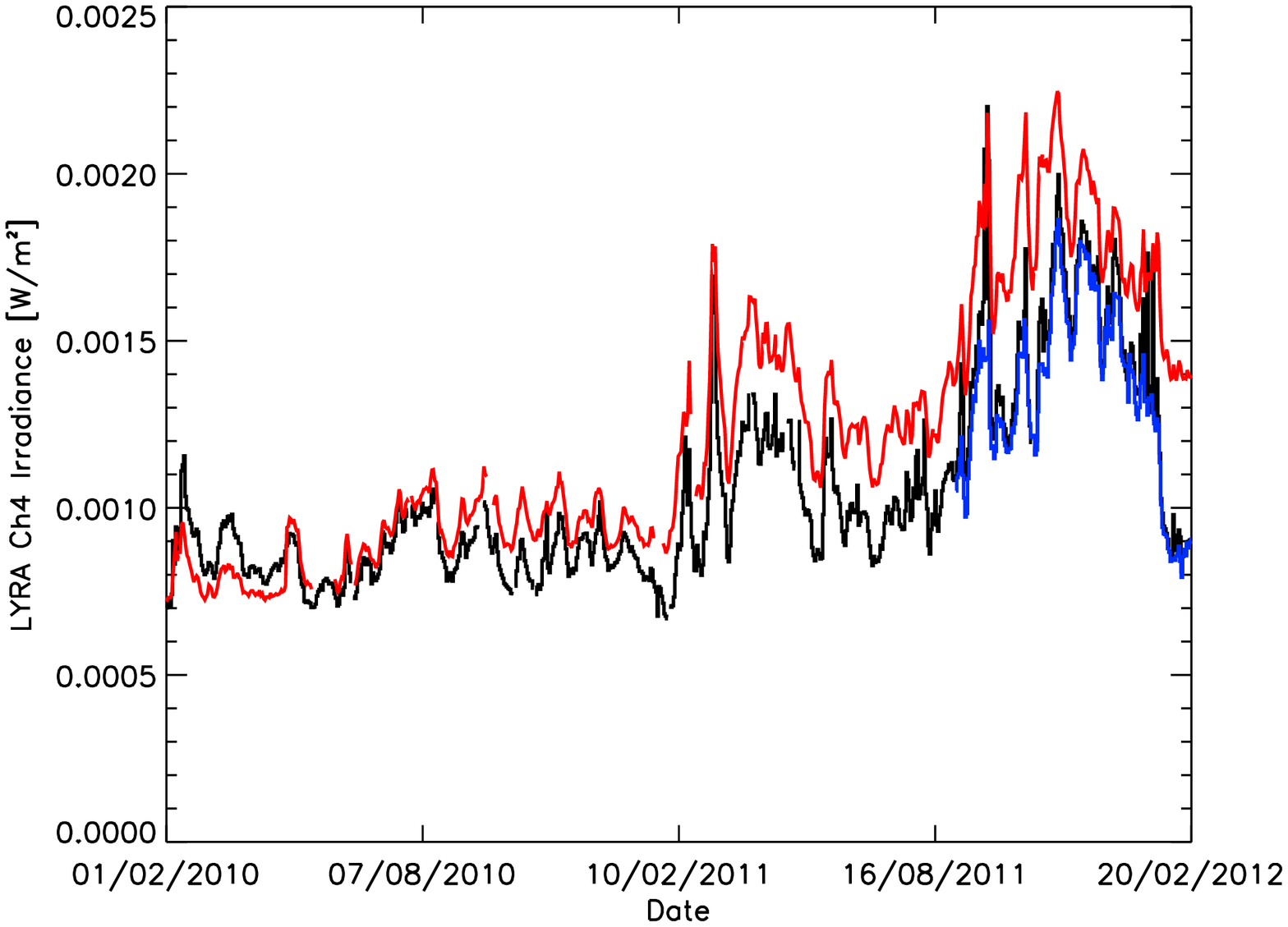}\includegraphics[width=0.49\textwidth,bb=40 190 600 600,clip=]{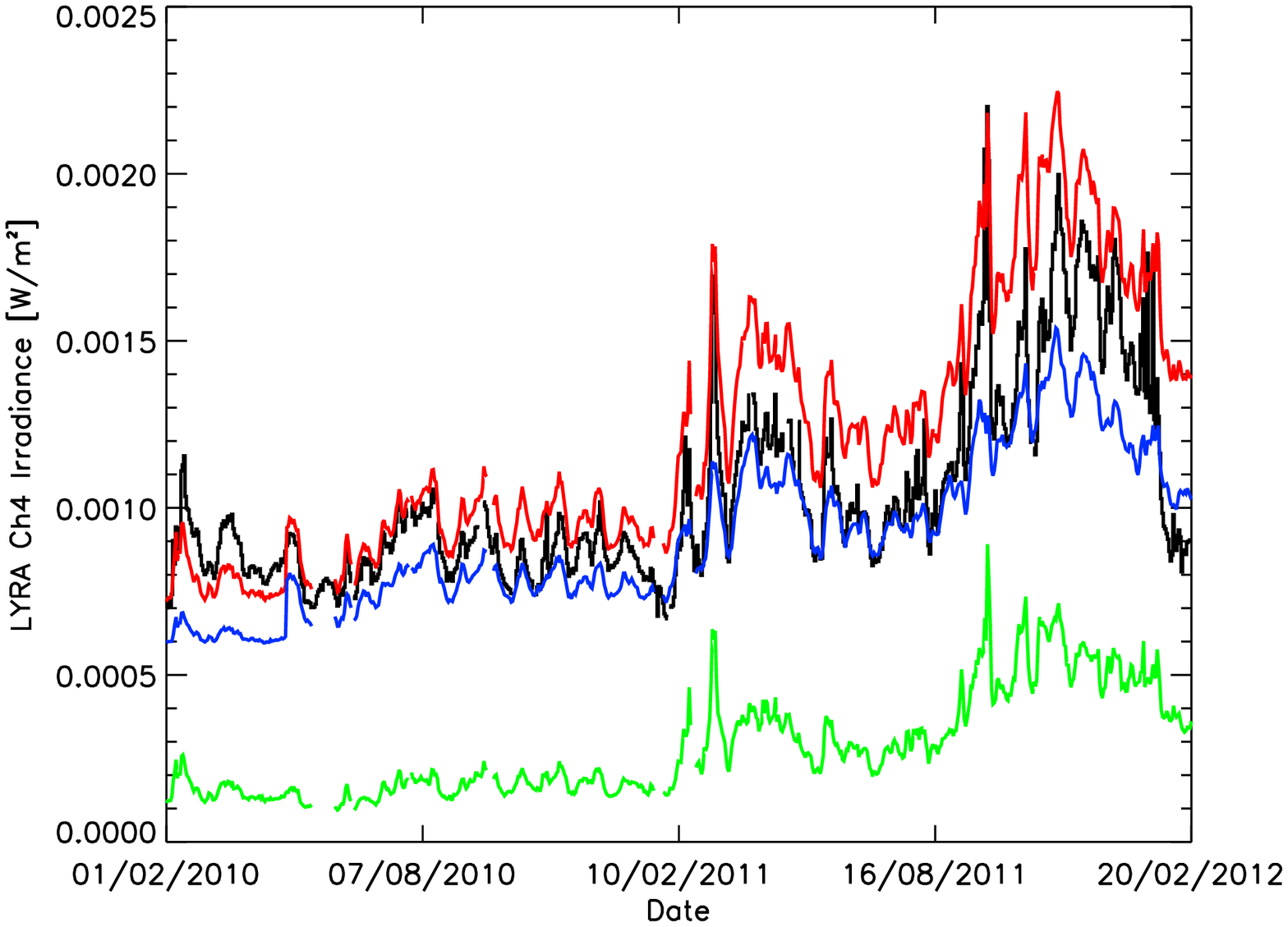}
             }
              \caption{\textbf{Daily value of LYRA channel 4 irradiance}. \emph{Left panel:} Lyra channel 4 time series (in black) and comparison with simulated response using TIMED/SEE and SDO/EVE (in red). The blue line shows the channel 4 daily minimum value. \emph{Right panel:} In black, the daily mean LYRA channel 4 irradiance; in red, the simulated value. Blue and green lines indicate the EUV and SXR contributions to the simulated channel 4, respectively.}
   \label{fig_ch4_simu}
   \end{figure}
The mean daily irradiances for LYRA channel 3 and 4 are shown in the next figures. In order to compare and validate these time series, we have simulated the LYRA signals based on the TIMED/SEE and SDO/EVE observations. We first built for each day a solar spectrum from 0.5nm to 190nm using the TIMED/SEE level 3 irradiance (version 10) and the SDO/EVE level 3 irradiance (version 2). The EVE data cover the period after 30 April 2010 and the wavelengths from 7.5nm to 64.5nm; the data were rebinned to the TIMED/SEE level 3 spectral resolution of 1nm. The TIMED/SEE irradiance shortward of 27nm (that we used only below 7.5nm after 30 April 2010) are deduced from the TIMED/SEE/XPS measurements: a solar spectrum model is scaled to match the XPS broadband data. Unfortunately, true daily spectra at these wavelengths are not available. For each day, the daily spectrum was multiplied with the LYRA channel 3 and 4 spectral response and the simulated irradiance was computed by applying the usual LYRA conversion from current to W/m$^{2}$. 

Fig.\ref{fig_ch4_simu} shows the LYRA channel 4 daily irradiance together with its simulation. Starting in autumn 2011, a daily minimum value of LYRA channel 3 and 4 has been retrieved by hand after the examination of the daily curve; this has been done for for the purpose of flare studies, but we use it here to validate the daily mean time series. Indeed, the agreement between both curves is excellent and shows the reliability of our daily value. The overall agreement between the daily values and the simulated ones, both on absolute scale and variations, is very good. We can note a difference in long term variation, i.e., the SDO/EVE and TIMED/SEE data have a larger increase with the rise of the solar cycle. Since both LYRA and EVE are young instruments, the degradation estimates are still  preliminary (version 2 in each case), and we expect the agreement to improve in the future. 

In order to better understand how the different wavelength contribute to the LYRA signals, the right panel of Fig.\ref{fig_ch4_simu} shows the contributions of the long ("EUV", above 5nm) and short ("SXR", below 5nm) wavelengths separately. The EUV dominates both the absolute value of the LYRA channel 4 signal and its variations. We also note how the SXR contribution increases when hot active regions and flares appear, see for example, just after 10 February 2011 and 16 August 2011. 
In both panel of Fig.\ref{fig_ch4_simu}, we can note a discrepancy on the irradiance level in the early days of 2012. The reason is unclear yet, but because the solar activity has significantly increased since August 2011, this could be attributed to a larger degradation of channel 4 at EUV wavelength, that is not yet corrected in the current version. 

%
 \begin{figure}[t]    
   \centerline{\includegraphics[width=0.49\textwidth,bb=40 190 600 600,clip=]{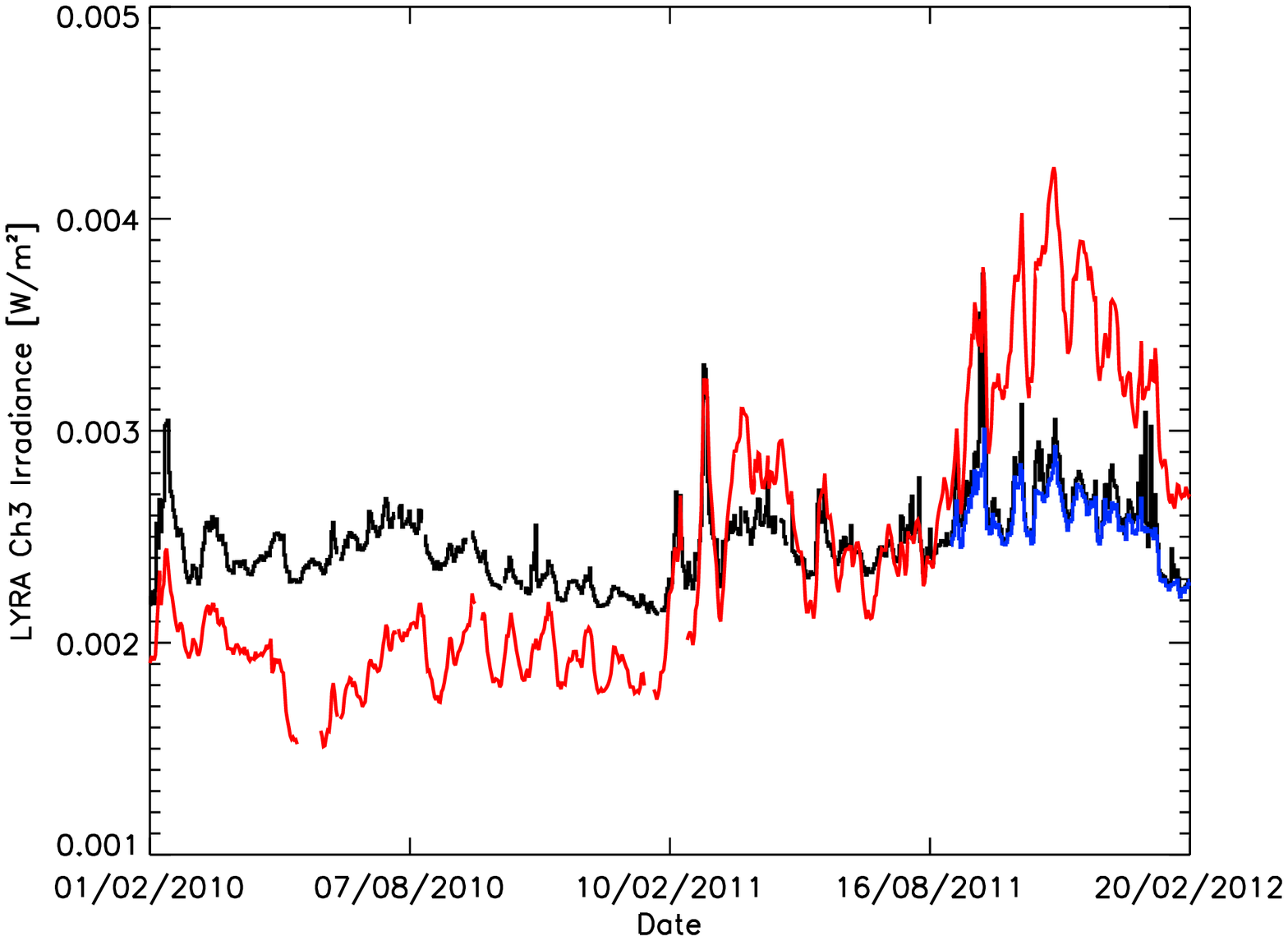}\includegraphics[width=0.49\textwidth,bb=40 190 600 600,clip=]{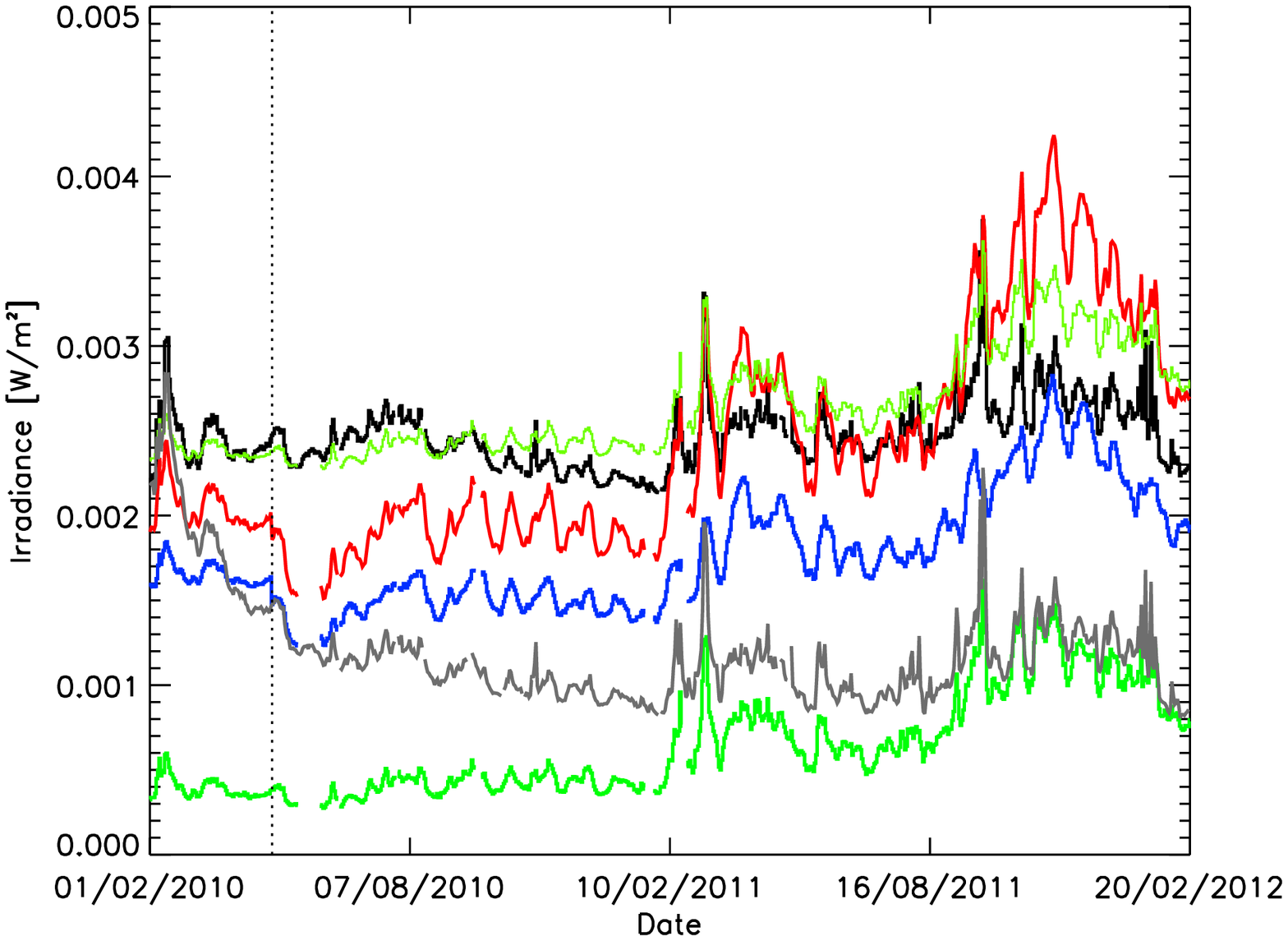}
             }
            \caption{\textbf{Daily value of LYRA channel 3 irradiance.} \emph{Left panel}: daily channel 3 irradiance (black) compared with the simulated response using TIMED/SEE and SDO/EVE (in red). The blue line shows the channel 3 daily minimum value. \emph{Right panel}: same as in the left panel for the black and red lines. Dark blue and green lines indicate the EUV and SXR contributions to the simulated channel 3, respectively. The thin light green line shows the SXR contribution scaled to the LYRA channel 3 level (by adding 2.10$^{-3}$W/m$^{2}$). The thick grey line shows LYRA channel 3 uncorrected for degradation. The start date for the use of the EVE data is indicated by the dotted vertical line. }
   \label{fig_ch3_simu}
   \end{figure}
%
%
%
The left panel of Fig.\ref{fig_ch3_simu} shows the LYRA channel 3 daily irradiance together with its simulation. Here again, the agreement between the daily mean channel 3 (automatically computed) and the daily minimum selected by hand is excellent and validates the reliability of the daily time series. There is a good agreement on the absolute scale for the daily values and their simulation but the simulated time-series has a much larger variability. As shown in Fig.\ref{fig_resp}, the SXR radiation, which is the most variable, accounts for about 20\% of the channel 3 signal. However, careful inspection of the simulated time series for the SXR ($\lambda <$ 17) and EUV ($\lambda \geqslant $17) contributions, as shown in the right panel of Fig.\ref{fig_ch3_simu}, reveals that the missing variability in LYRA channel 3 is caused by the absorption of its EUV component. Indeed, the simulated SXR contribution to channel 3 has a very similar variability as the observed channel 3 irradiance after about February 2011. Furthermore, the channel 3 irradiance not corrected for degradation is now at the level of (and following) its SXR component. In other words the EUV contribution to channel 3 has almost completely disappeared as a consequence of degradation. This indicates that the current degradation correction corrects for the absolute irradiance level but not for the amplitude of the variability. We discuss this aspect in detail in the next section. 
\subsection{Degradation of LYRA channel 3}
 \begin{figure}[t]    
   \centerline{\includegraphics[width=0.49\textwidth,bb=40 190 600 600,clip=]{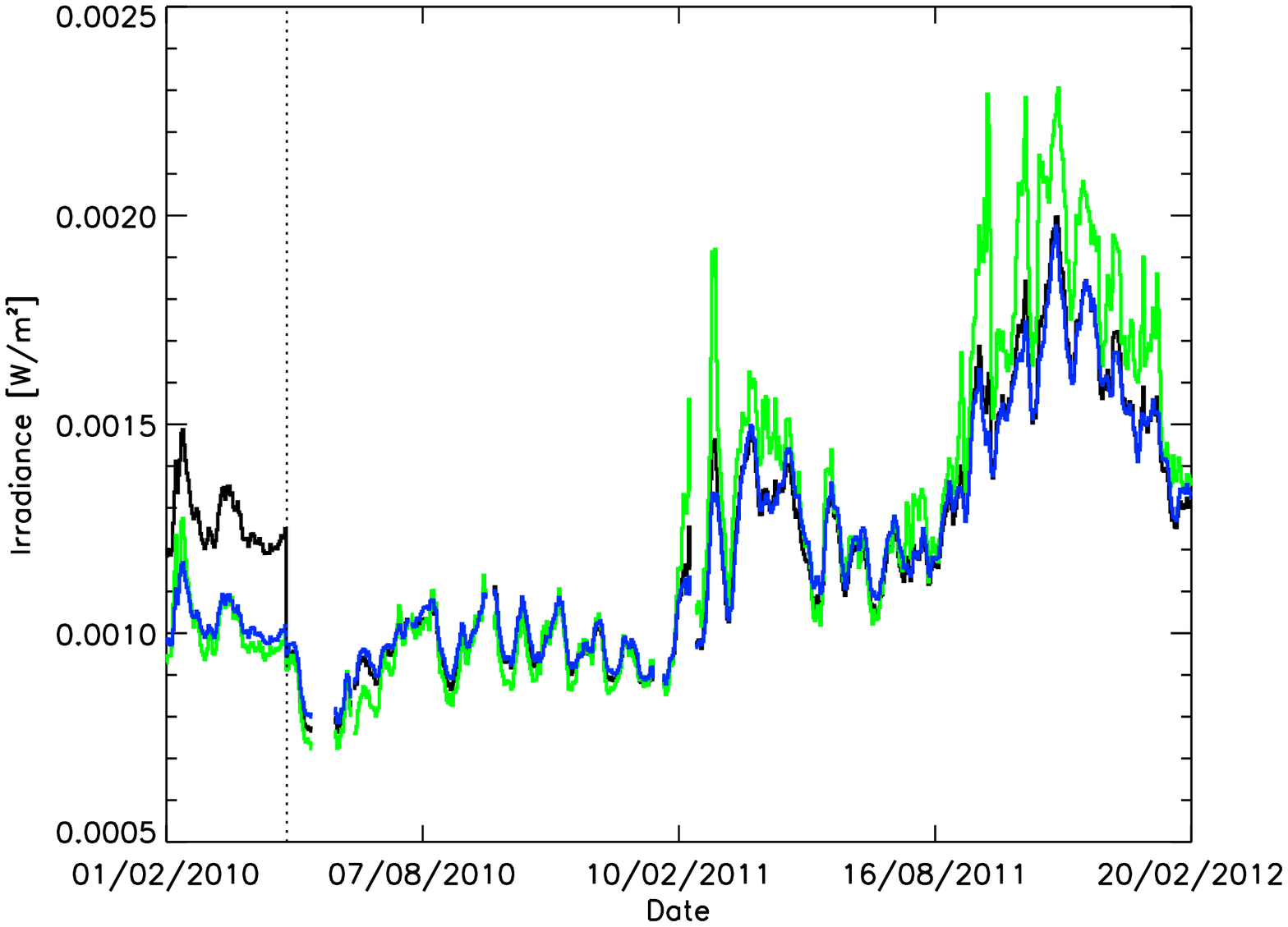}
   \includegraphics[width=0.49\textwidth,bb=40 190 600 600,clip=]{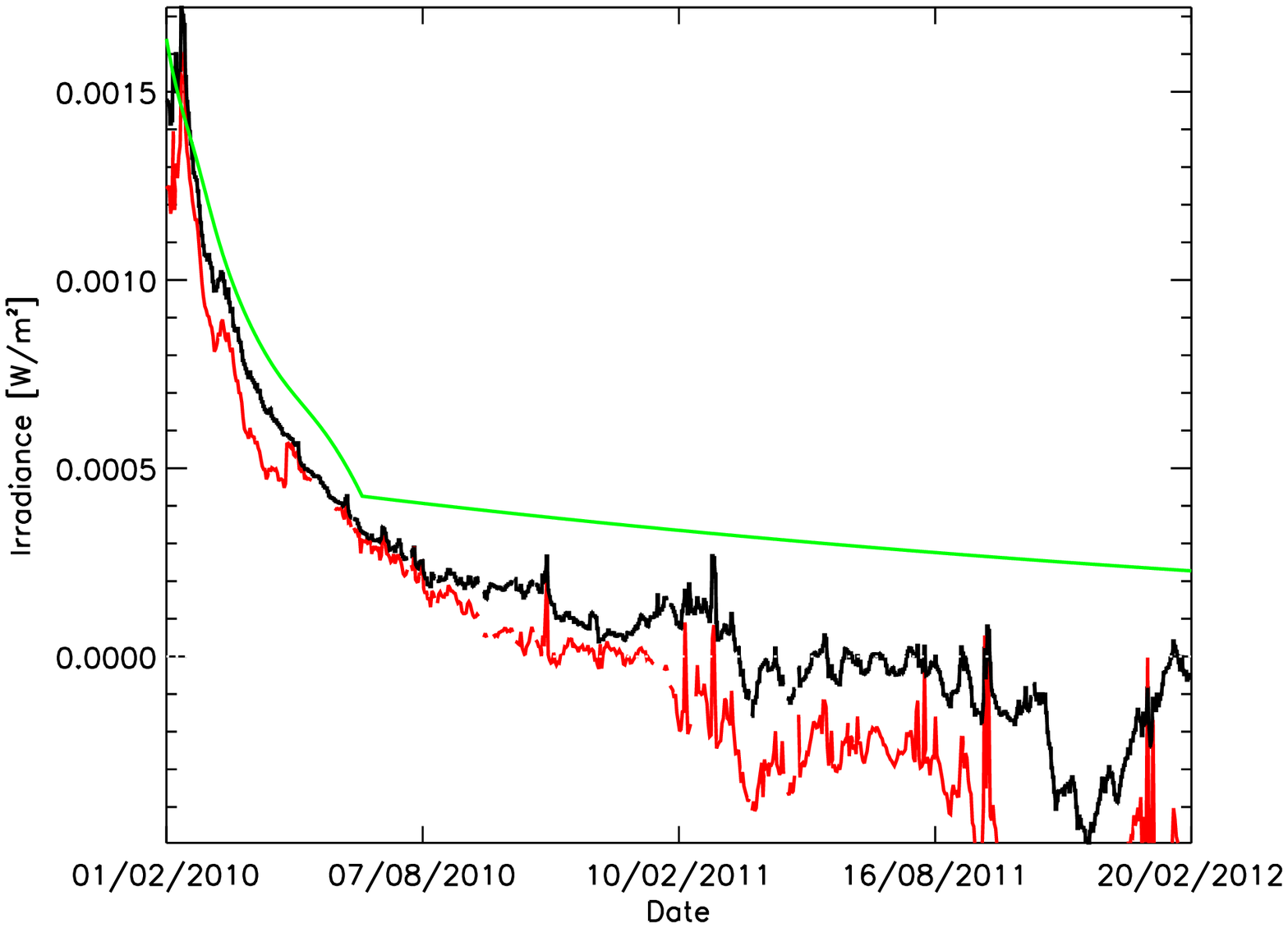}
             }
              \caption{Left panel: In black, the difference between the simulated channels 3 and 4; in green, the simulated contribution to the channel 3 signal from wavelengths shorter than 19nm. In blue, the simulated contribution for wavelengths longer than 19nm. The start date for the use of the EVE data is indicated by the dotted vertical line. Note that there is the same contribution to channel 3 from wavelengths shorter than 19nm as from wavelengths longer than 19nm to channel 3. Right Panel: the degradation of channel 3 response above 19nm based on channel 4 (black) and TIMED/SEE and SDO/EVE (red, see text for explanations).The green line shows the current estimate of the degradation}
   \label{fig_ch34diff_simu}
   \end{figure}
Fig.\ref{fig_ch3_simu} shows that the EUV part of the channel 3 signal has disappeared as a consequence of degradation, while the SXR contribution has not, or is slightly degraded. This is confirmed by the much smaller degradation of channel 4 (whose pass band stops at 20nm, see sec.\ref{sec_lyracal}). Ideally, one would wish to distinguish the different behaviors of the SXR and EUV contribution of channel 3   without the need to refer to another instrument, which is itself subject to degradation. One way to do this is provided by the relatively same shape of the channel 3 and channel 4 spectral responses at the short wavelengths (see Fig.\ref{fig_resp}): by subtracting the channel 4 signal from the channel 3 signal, one can hope to be left with the EUV contribution to channel 3 only. The left panel of Fig.\ref{fig_ch34diff_simu} shows the simulated (channel 3 - channel 4) signal, together with the simulated contribution to channel 3 of wavelengths shortward (in green) and longward (in blue) of 19nm. It occurs that the signal coming from wavelengths longward of 19nm fits exactly the difference signal , so that we can use this difference to monitor the behavior of the EUV contribution to channel 3. From now on, we rename the EUV component of channel 3 as the contribution coming from wavelengths above 19nm, where we assume that the degradation is concentrated. Note that this looks reasonable since SWAP, the EUV imager onboard PROAB2 which has also an Aluminum entrance filter but which observes at 17nm (using coated mirrors), has not shown significant degradation so far (D. Seaton, private communication). Additionally, it is interesting to note that dividing the channel 3 signal in wavelengths shorter and longer than 19nm fortuitously leads to two contributions of the same amplitude. The drop of the simulated signal after 30 April 2010 is caused by the use of SDO/EVE data instead of TIMED/SEE data for wavelengths between 7 and 64nm after that date (SDO/EVE and TIMED/SEE are not yet inter-calibrated).\\

The right panel of Fig.\ref{fig_ch34diff_simu} shows the degraded EUV component ($\lambda > $19nm) of channel 3, computed by removing the SXR component from the degraded channel 3 signal. This has been done in two ways: first by subtracting (the degradation corrected) channel 4 to (the non-corrected) channel 3 (thus, using only LYRA data, black curve), and second by subtracting the simulated contribution of wavelengths shorter than 19nm to channel 3. The two estimates are very similar. By comparison, we show the degradation of channel 3 as a whole as it is estimated now. The figure suggests that the degradation has been partially hidden by the growing SXR and, therefore, might currently be slightly underestimated after August 2010; however, adding the degradation as it is deduced using channel 4 will not solve the too small variability of channel 3. Instead, one should rather attempt to correct the degradation by multiplication rather than addition, i.e., to look at the ratio of the degraded signal to the non-degraded signal $  \textrm{L}^{deg}_{Ch3/EUV}(t) / \textrm{L}_{Ch3/EUV}(t)$ (where ch3 refers to channel 3, deg indicates the non-corrected observations, and ch3/euv indicates the EUV component). 
%
 \begin{figure}[t]    
   \centerline{ \includegraphics[width=0.49\textwidth,bb=40 190 580 600,clip=]{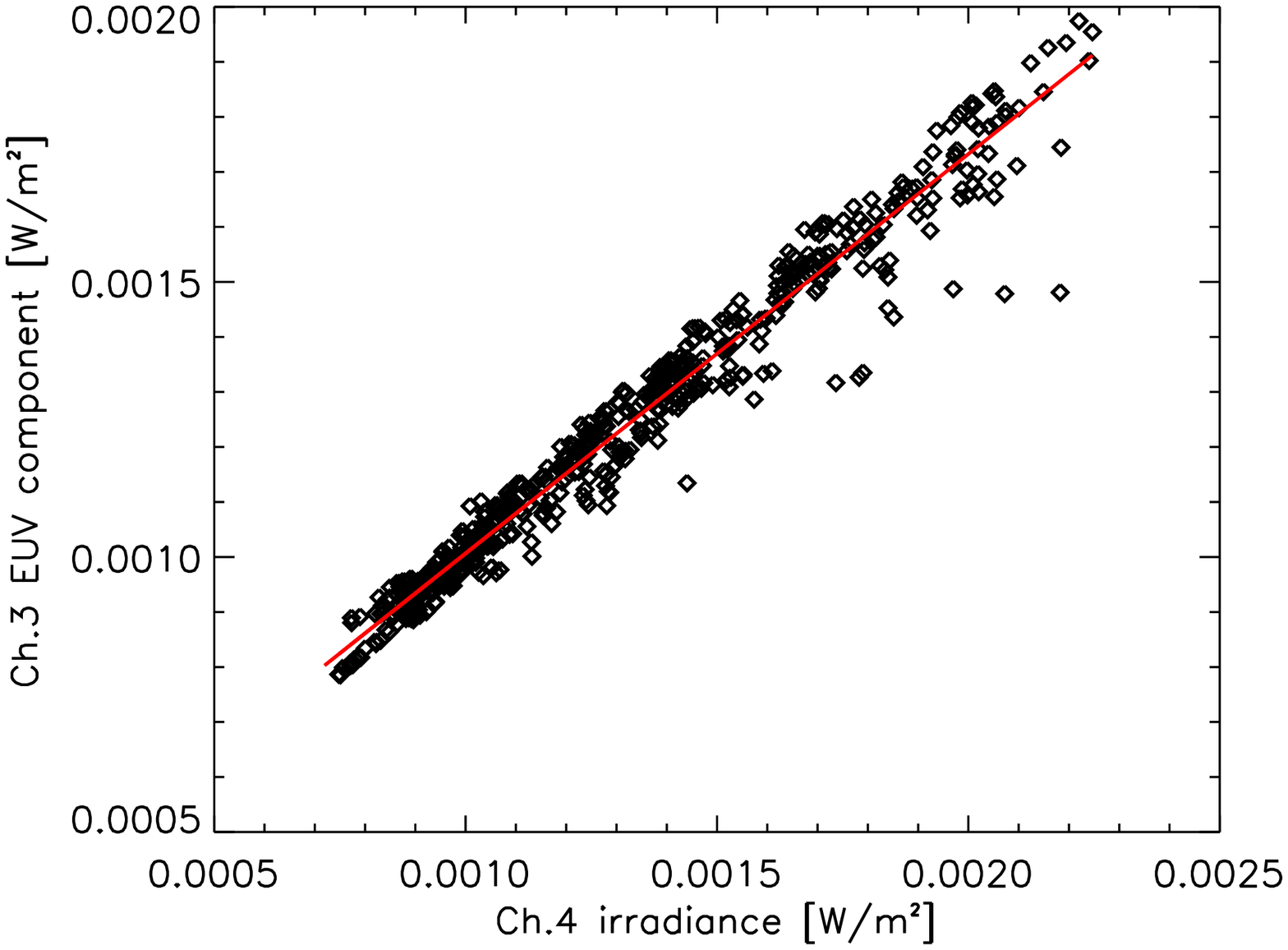}
             \includegraphics[width=0.49\textwidth,bb=50 190 580 600,clip=]{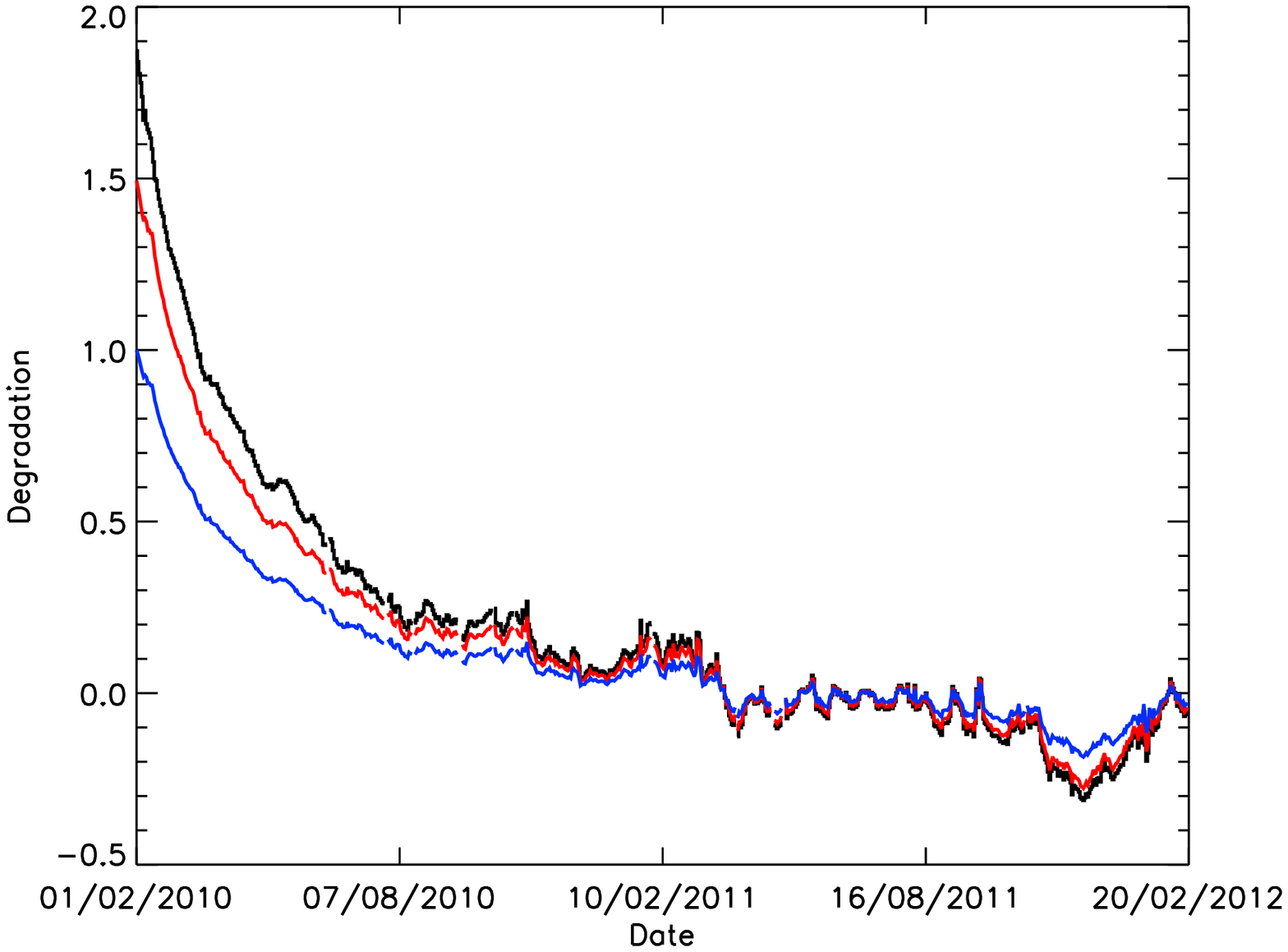}
             }
              \caption{\emph{Left panel}: Simulated EUV component of channel 3 vs simulated channel 4. \emph{Right panel}: The ratio $   \textrm{L}^{deg}_{Ch3/EUV}(t) / \textrm{L}_{Ch3/EUV}^{simu}(t)$ showing the (multiplicative) degradation of the EUV component. In black, $ \textrm{L}^{simu}_{Ch3/EUV}(t)$ has been determined to match the EVE data; in red, to match the TIMED/SEE data; In blue, the normalized curve. }
   \label{fig_ch3_EUV_vs_ch4}
   \end{figure}

The problem here is how to estimate the non-degraded behavior of the EUV component of channel 3, $\textrm{L}_{Ch3/EUV}(t)$. To demonstrate the need of a multiplicative correction to the degradation, and still based on LYRA observations as much as possible, we assume that this non-degraded component can be derived by simply scaling channel 4, assuming that channel 4 is correctly corrected. The left panel of Fig.\ref{fig_ch3_EUV_vs_ch4} shows that the relation L$_{Ch3/EUV}^{simu}$=2.8 10$^{-3}$+0.726 L$_{Ch4}$, derived from simulated values, is quite robust (note this relation is based with data from EVE only; with data from TIMED/SEE only, the relation becomes L$_{Ch3/EUV}^{simu}$=3.10$^{-3}$+0.726 L$_{Ch4}$). 

The right panel of Fig.\ref{fig_ch3_EUV_vs_ch4} shows the ratio $  \textrm{L}^{deg}_{Ch3/EUV}(t) / \textrm{L}_{Ch3/EUV}^{simu}(t)$, where $\textrm{L}_{Ch3/EUV}^{simu}(t)$ is now computed using the relation above and the actual LYRA channel 4. The fact that this ratio is greater than one reflects the difference in calibration and in degradation between instruments; if we used SDO/EVE and/or TIMED/SEE to compute $\textrm{L}_{Ch3/EUV}^{simu}(t)$, we would simply match the respective irradiance level of the simulated EUV component. In order to implement an independent correction, we normalize this ratio to unity. The normalized degradation curve (shown in blue in Fig.\ref{fig_ch3_EUV_vs_ch4}) for the EUV component is then,
$$ \textrm{D}_{Ch3/EUV}  = \frac{ \textrm{L}_{Ch3}^{deg} - \textrm{L}_{Ch4} }{ 2.8 10^{-3}+0.726 L_{Ch4} }  \times      
 \frac{2.8 10^{-3}+0.726 L_{Ch4}(t=0)}{\textrm{L}_{Ch3}^{deg}(t=0)-\textrm{L}_{Ch4}(t=0)} $$
$$ \textrm{D}_{Ch3/EUV}  = \frac{ \textrm{L}_{Ch3}^{deg} - \textrm{L}_{Ch4} }{ \textrm{L}_{Ch3}^{simu} }  \times      
 \frac{\textrm{L}_{Ch3}^{simu}(t=0)}{\textrm{L}_{Ch3}^{deg}(t=0)-\textrm{L}_{Ch4}(t=0)} $$
 , the corrected EUV component is determined by 
$$\textrm{L}_{Ch3/EUV}  = \frac{(\textrm{L}_{Ch3}^{deg}-\textrm{L}_{Ch4}) }{ \textrm{D}_{Ch3/EUV}}$$
, and the new whole channel 3 by
$$\textrm{L}_{Ch3}  = \textrm{L}_{Ch3/EUV}+\textrm{L}_{Ch4}$$

 \begin{figure}[t]    
   \centerline{\includegraphics[width=0.7\textwidth,bb=1 190 600 600,clip=]{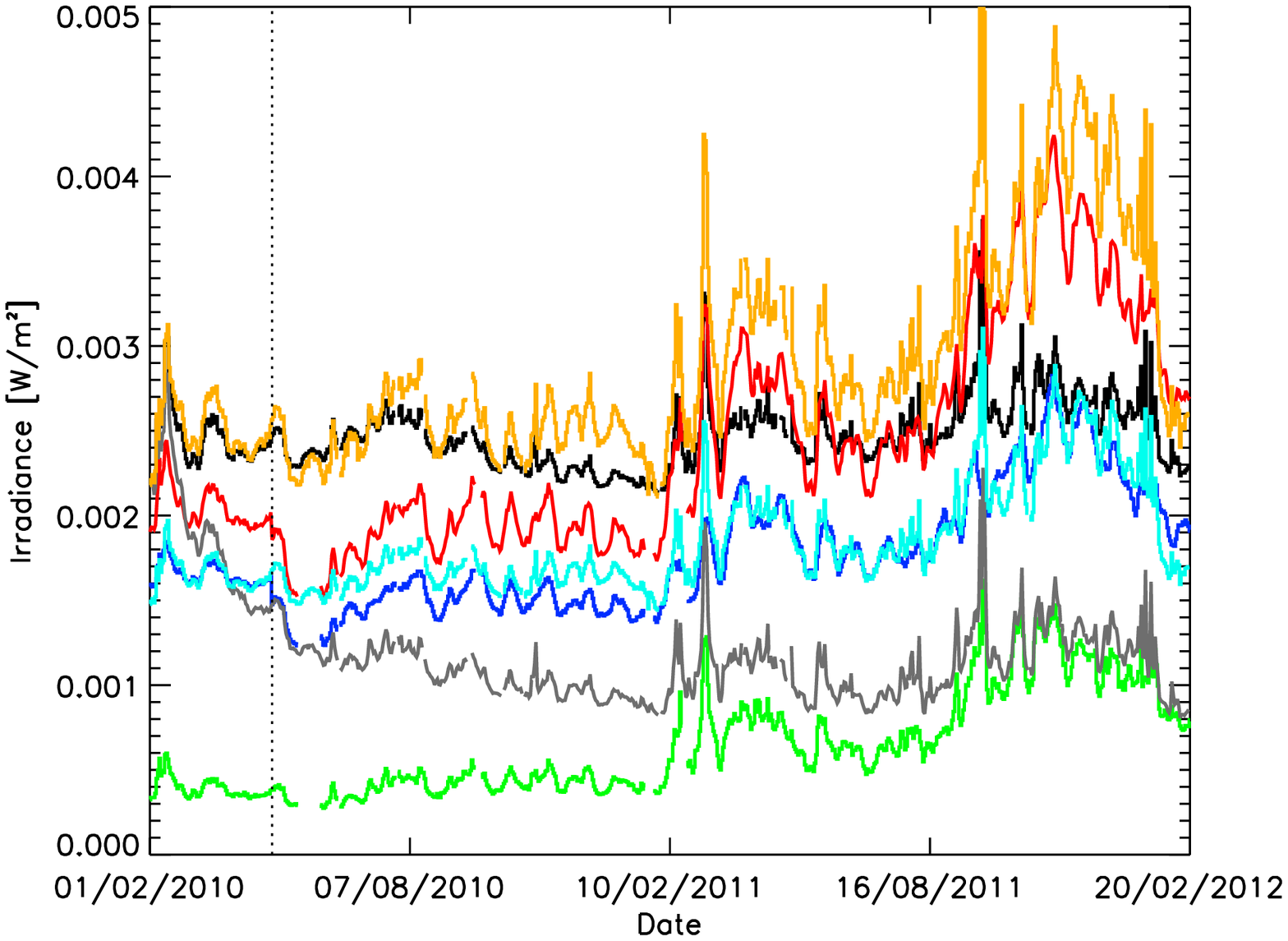}
             }
              \caption{\textbf{Correction of the degradation of the EUV component of LYRA channel 3}. Same colors as in Fig.\ref{fig_ch3_simu}: the red, dark blue and green lines show the simulated LYRA channel 3 and its EUV ($\lambda >$ 17nm) and SXR components ($\lambda <$ 17nm), respectively; the  black and grey lines show the actual LYRA channel 3, corrected (by addition) and not corrected for degradation. The light blue curve shows the EUV component of channel 3 corrected for degradation by multiplication. The orange curve shows the whole LYRA channel 3 irradiance, with its EUV component corrected for degradation by multiplication. The start date for the use of the EVE data is indicated by the dotted vertical line.}
   \label{fig_ch3_mcorr}
   \end{figure}
Fig.\ref{fig_ch3_mcorr} shows the result of applying this multiplicative correction. The corrected EUV component agrees well with the simulated one, both showing the same amount of variability with the solar cycle and solar rotation. The discrepancy at the end of 2012 can be explained if we assume that the degradation of EVE has been underestimated at the beginning of the SDO mission. 

The newly corrected LYRA channel 3 predicts a higher flux than the one simulated from SDO/EVE and TIMED/SEE. This probably comes from differences in the degradation correction of the instruments, in particular of TIMED/SEE for the SXR part. Howvere, we note that the overall variability is now well reproduced, which demonstrates the needs for a multiplicative correction (maybe together with an additive term) in order to correct for the degradation. Of course, as the actual EUV contribution to channel 3 decreases, the noise increases in the corrected time series. We note that the increase of channel 3 irradiance between February 2010 and November 2011 is about a factor two, which agrees with the data from SDO/EVE and TIMED/SEE. Its EUV component has increased slightly less, by about a factor 1.9, the difference being explained by the SXR contribution to channel 3.
%

%

%
\section{Conclusions\label{sec_conclu}}
In this paper, we have presented methods to correct for several artifacts present in the LYRA data in order to build daily time series over the mission lifetime. This long term time series has been compared with the results of the SDO/EVE and TIMED/SEE instrument. It has been shown that the LYRA channel 4 is in good agreement with the SDO/EVE and TIMED/SEE observations. A slightly larger trend with the rising solar cycle has been detected for the SDO/EVE irradiance. LYRA channel 3 has been shown to be at a similar level as predicted by SDO/EVE and TIMED/SEE, but with a much smaller variability. We have shown that this is due to the strong degradation of the longer wavelengths included in its pass band, above approximately 19nm and that we have termed the EUV component. Presently, channel 3 is clearly dominated by wavelength below 17nm. We have shown that the degradation of the EUV component of the LYRA channel 3 can be corrected by multiplication with a degradation term, and that this term can be  deduced using channel 4 and an empirical relation between the two channels. We plan to improve on the degradation correction of channel 3 in the future. Finally, both the LYRA channel 3 and 4 irradiance has increased between February 2010 and November 2011 by approximately a factor of two, channel 4 having a slightly larger increase than channel 3. These results are  useful for analyzing and/or modeling other space weather relevant variables, like the TEC and thermosphere densities. 

%
\section{Acknowledgements}
LYRA is a project of the Centre Spatial de Li\`ege, the Physikalisch- Meteorologisches Observatorium Davos and the Royal Observatory of Belgium funded by the Belgian Federal Science Policy Office (BELSPO) and by the Swiss Bundesamt f\"ur Bildung und Wissenschaft. M.D. and I.E.D acknowledge the support from the Belgian Federal Science Policy Office through the ESA-PRODEX programme. This work has received funding from the European Community's Seventh Framework Programme (FP7/2007-2013) under the grant agreement n¡ 261948 (ATMOP project, www.atmop.eu). MK acknowledges Stefania and Francesco for logistic support. 
%
%
\bibliographystyle{plainnat} 
\bibliography{Kretzschmar_LyraLongTerm} 

\end{document}